\documentclass[twocolumn]{jetc20132e}
\usepackage{graphics,amsmath}

\newcommand{\pd}[2]{\frac{\partial #1}{\partial #2}}

\def\d{{\rm d}} 

\newcommand{\be}{\begin{equation}}
\newcommand{\ee}[1]{\label{#1} \end{equation}}
\newcommand{\ba}{\begin{eqnarray}}
\newcommand{\ea}[1]{\label{#1} \end{eqnarray}}
\newcommand{\nl}{\nonumber \\}
\newcommand{\re}[1]{(\ref{#1})}

\def\re#1{(\ref{#1})}   

\title{Dissipation flow-frames: particle, energy, thermometer}

\author{P. V\'an$^{1,2,3}$ and T.S. Bir\'o$^{1}$ 
\affiliation{ 
    $^1$Dept. of Theoretical Physics, Wigner Research Centre for Physics, Institute for Particle and Nuclear Physics, \\  
      H-1525 Budapest, Konkoly Thege Mikl\'os \'ut 29-33, Hungary; \\
    $^2$Dept. of Energy Engineering, Budapest Univ. of Technology and Economics,\\
      H-1111, Budapest, M\H uegyetem rkp. 3-9,  Hungary; \\
    $^3$Montavid Thermodynamic Research Group \\
Email: van.peter@wigner.mta.hu}}

\date{
{\small 
We associate the following physical co-mover conditions of to different frame choices: i) Eckart: particle flow, ii) Landau-Lifshitz: energy flow, iii) J\"uttner: moving thermometer frame. The role of fixing a flow-frame is analysed with respect to local equilibrium concentrating on dissipative currents and forces  in single component relativistic fluids. The special role of a ''J\"uttner frame'' is explored and contrasted to the more common Eckart and Landau-Lifshitz choices. 
}}

\begin{document}

\maketitle

\section{Introduction}

In dissipative theories of relativistic fluids we deal with four fundamental questions. 

The first considers \textit{causality.} Only divergence type theories are, in general, causal because there the symmetric hyperbolicity of the system of nonlinear evolution equations is established by construction \cite{Ger01m,Ger95a,GerLin90a,GerLin91a,LiuAta86a,PerCal10a}. The weaker version of causality requires for the symmetric hyperbolicity only for the linearized equations, and allows for characteristic speeds less than the speed of light \cite{ShiEta10a}. This weak causality was studied in the Israel-Stewart theory; numerous resulting inequalities are given in  \cite{HisLin83a}. From a physical point of view the causality of theories with parabolic differential equations should also be possible. In this case the validity of the continuum description is restricted by the characteristic maximal speeds \cite{Fic92a,KosLiu00a,Cim04a,VanBir08a}. A necessary condition for this type of restrictions requires the damping of the perturbations, equivalent to the the linear stability of the theory \cite{DenAta08a1}. 

The second question deals with \textit{generic stability.} Generic stability is the linear stability of the homogeneous equilibrium solutions. The simplest relativistic generalization of the nonrelativistic Fourier-Navier-Stokes equations was proved to be unstable by Hiscock and Lindblom \cite{HisLin85a}. In the sequel they formulated mathematical conditions of  generic stability of the Israel-Stewart theory \cite{HisLin83a}
specified to the Eckart frame. However, due to the overwhelming complexity of these conditions they are not connected to reasonable properties of equations of state or transport coefficients. Since then several different propositions arose suggesting a first order theory, mostly motivated  by the restoration of the generic stability  \cite{TsuKun08a,TsuKun11a,VanBir08a,Van09a,VanBir12a,GarEta09a,Osa10a,Osa12a}.    

The third question is the correct \textit{distinction between ideal and dissipative fluids}, especially  in a relativistic context. It is customary to assume that perfect, nondissipative fluids are characterized by a special form of the energy-momentum tensor and the particle current density  vector \cite{GroAta80b,Cse94b}. On the other hand physical dissipation is accompanied by nonzero entropy production. From this point of view there is a more extended family of perfect fluids beyond the customarily treated ones \cite{Van11p}. These distinctions are technically addressed by the so called matching conditions \cite{IsrSte79a1,Osa10a,Osa12a,MonHir09a}.

Finally the proper choice of \textit{flow-frames} continues to be an unsettled question \cite{TsuKun11a}. One generally believes that in relativistic fluids the flow field \(u^{a}\) can be chosen arbitrarily, since it is a somewhat vaguely defined physical property, belonging to the flow of volatile quantities, once the energy, once the conserved charge. In this situation it is customary to fix the flow either to  the motion of particles (Eckart frame) \cite{Eck40a3}, or that of the energy density (Landau-Lifshitz frame) \cite{LanLif59b}. The flow fixing determines a continuous set of local rest frames in the fluid:\ we shall refer to the different choices of  fixing the flow as \textit{flow-frames or frames}. Contrary to the belief in a free choice of the flow-frame we point out that this may not be completely arbitrary, as one associates a given physical content of the dissipation to each. Further choices than the two classical ones should be preferred by demanding a given form of local Gibbs relations.

In this paper we present the general flow-frame, the separation of perfect and dissipative parts of  energy-momentum and particle number current density and  their relation to  generic stability. The key theoretical aspect connecting these problems is relativistic thermodynamics. Our most important observation is that the usual assumption of kinetic equilibrium  by introducing the velocity field parallel to the local thermometer and Lagrange multiplier field \(\beta^a\) also appearing in the collision invariant \(\psi= \alpha+\beta_ak^a\), already acts as a flow-frame fixing. This choice we tag as thermometer frame or J\"uttner frame,  distinguishing from the Eckart, Landau-Lifshitz and other conventions.

\section{General one component dissipative relativistic fluids}

In this paper the Lorentz form is given as $g^{ab} = diag(1, −1, −1, −1)$ and all indexes $a,b,c,...$ run over
$0, 1, 2, 3$. We use natural units, $h = k = c = 1$.

A single component fluid is characterized by the particle number four-vector \(N^{a}\) and
the symmetric energy-momentum density tensor \(T^{ab}\). The velocity field of the fluid, the flow-frame $u^a$, introduces a  local rest frame and  the basic fields \(N^a\) and \(T^{ab}\) can be expanded by their local rest frame components parallel and perpendicular to the flow:
\ba
N^{a}&=&nu^a+j^a, \label{N_sp}\\ 
T^{ab}&=&eu^au^b+q^au^b+u^aq^b+P^{ab}.
\ea{T_sp}
Here \(n\) is the flow-frame particle number density, \(j^{a}\) is in this frame the non-convective particle number current density, \(e\) is the energy density, \(q^{a}\) is the momentum density and \(P^{ab}\) is the pressure tensor. These components are flow-frame dependent, in particular \(j^{a}u_a=0\), \(q^au_a=0\) and \(P^{ab}u_b=0\). Introducing the substantial time derivative $\frac{d}{dt}:= u^a\partial_a$ denoted by and overdot, the balances of  the particle current density and energy-momentum are expressed by the local rest frame quantities:
\ba
\partial_a N^a &=& \dot n + n\partial_au^a + \partial_aj^a=0,\label{pnumc_bal}\\
\partial_b T^{ab} &=&  \dot e u^a +  e\dot u^a + eu^a\partial_bu^b +\dot q^a 
    +q^a\partial_b u^b +    \nl
        &&     u^a\partial_b q^b+q^b\partial_b u^a +\partial_bP^{ab}=0^{a}. 
\ea{emomc_bal}

The energy and momentum balances are the time and spacelike components of the energy-momentum balance projected in the flow-frame: 
\ba
u_a\partial_b T^{ab} &=&  \dot e  +e\partial_bu^b +u_a\dot q^a +\partial_b q^b -P^{ab} \partial_bu_{a}= 0, \label{e_bal}\\
\Delta^a_c\partial_b T^{cb} &=&   e\dot u^a +\Delta^a_b\dot q^b +q^a\partial_b u^b + q^b\partial_b u^a +\Delta^a_c\partial_bP^{cb} =0^a. 
\ea{momc_bal}

The frame related quantities are important in the separation of the ideal and dissipative parts of the basic fields. This separation is best performed by analyzing the thermodynamical relations. In order to achieve this one postulates the existence of an additional vector field, the entropy current as a function of the basic fields \(S^{a}(N^a, T^{ab})\). It must not decrease by fulfilling the condition of the balances \re{pnumc_bal} and \re{emomc_bal}. That conditional   inequality can be best represented by introducing the Lagrange-Farkas multipliers\footnote{Lagrange multipliers are introduced for conditional extrema.  For conditional inequalities Gyula Farkas suggested analogous quantities and proved the corresponding theorem of linear algebra, called Farkas' lemma  \cite{Far894a,Min896b,Sch98b}.} \(\alpha\) and \(\beta^a\), respectively:\ 
\be
\Sigma := \partial_aS^a + \alpha \partial_aN^a - \beta_b\partial_aT^{ba} \geq 0.
\ee{entrcond}
The left hand side of this inequality shows, that the definition of the entropy production is done before specifying the flow-frame. However, the separation of ideal and dissipative parts of basic physical quantities, is a consequence to the choice of that flow-frame. Citing the authors of \cite{TsuKun12a}, when arguing about the uniqueness of the Landau-Lifshitz frame ``The uniqueness of the energy frame comes from ... the physical assumption that the dissipative effect comes from only the spatial inhomogeneity.``. However, what is spacelike is a frame dependent question and one hopes only that physical systems may reveal by their internal dynamics a physical ground for such a separation. A possible candidate for this separation can be the thermometer vector, $\beta^a$, reconstructable from observations of a multiparticle spectra stemming from a relativistic fluid.

\section{Thermodynamics of relativistic fluids -- equilibrium}

The concept of perfect fluids deals with the absence of dissipation, the entropy production vanishes:
\be
\Sigma_0 = \partial_aS_0^a + \alpha_{} \partial_aN^a_0 - \beta_b\partial_aT_{0}^{ba} = 0.
\ee{kin_pGrel}

The equilibrium entropy density \(S_0^a\) is connected to the equilibrium particle number density \(N_0^a\) and equilibrium energy-momentum density \(T_0^{ab}\) by the following definition of the isotropic pressure:
\be
   p_0\beta^a = S_0^a+\alpha N_0^a-\beta_bT_0^{ab}.
\ee{eqrel}

Standard kinetic theory definitions and calculations satisfy the above expressions. Then \(\alpha\) and \(\beta^a\) are coefficients in the collision invariant of the equilibrium distribution function, $\psi = \alpha+\beta_ak^a$, and the pressure is that of an ideal gas \(p_0=n_0T\) .

Kinetic theory describes a perfect fluid by the detailed balance requirement. Out of equilibrium dissipation can occur. In a dissipative fluid all physical quantities in principle deviate from their local equilibrium values. There also may exist non-dissipative currents (presumably driven by non-dissipating forces, like the Lorentz force in magnetic fields). The thermodynamic approach aims at  the separation of dissipative and non-dissipative local currents, in order to ensure the positivity of the expression \re{entrcond}. Physical freedom in the choice of a flow-frame should be restricted to different handlings of non-dissipative currents.

It is natural to introduce the \textit{J\"uttner frame} \(u^a_J\) defined by the direction of \(\beta^a\) (thermometer motion): 
\be
{u^a_J}=\frac{\beta^a}{\sqrt{\|\beta^a\beta_a\|}}.
\ee{Jflow} 

In that frame the equilibrium fields are decomposed as:
\ba
N^a_0 &=& n_J u^a_J, \label{eqnbal}\\
T^{ab}_0 &=& e_Ju^a_Ju^b_J-p\Delta^{ab}_J,\label{eqTbal}\\
S^a_0 &=& (\beta_{J} h_J-\alpha n_J)u^a_J,
\ea{eqsbal}
where \(h_J=e_J+p_0\) is the equilibrium enthalpy density in the J\"uttner frame, 
and \(\beta_{J} = \beta_a u^a_J = 1/T_J\) is the reciprocal J\"uttner temperature. \(\alpha\), \(\beta_a\) and \(p_0\) do not carry a frame index, because they are introduced before specifying the flow-frame. On the other hand the representations \re{eqnbal}-\re{eqsbal} are frame dependent.  In case of a general flow-frame \(u^a\), that is not parallel to \(\beta^a \), one can characterize this difference by introducing $w^a = \beta^a/(\beta^bu_b)-u^a$. Then \(w^a\) is orthogonal to \(u^a\) (\(w^au_a = 0\)) and  spacelike ($w_aw^a=-w^2$). The Lagrange multiplier four-vector, $\beta^a$, can be splitted as
\be
\beta^a = \beta_{J} u^a_J = \beta(u^a+w^a), 
\ee{genbeta}
where \(\beta = \beta^au_a\) is the reciprocal temperature in a general frame defined by \(u^a\). The equilibrium  fields in this frame are given as
\ba
N^a_0 &=& n_0u^{a} + j_{0}^a, \label{geqnbal}\\
T_0^{ab} &=& e _0u^{a} u^b +q_{0}^{a}u^b+q_{0}^bu^a - p\Delta^{ab}+\frac{q_{0}^{a}q_{0}^{b}}{h_{0}},\label{geqTbal}\\
S^a_0 &=& (\beta h_0+\beta w_bq^b_0-\alpha n_0)u^a+\beta q^a_0-\alpha j_0^a.
\ea{geqsbal}

Here \(\beta=\beta_{J}/\sqrt{1-w^2} \), \(\ n_0 ={n_J}/{\sqrt{1-w^2}}\), \(\ e_0 = {(e_J+p w^2)}/{(1-w^2)}\), \(\alpha\) and \(p_0\) does not change, \(j_{0}^{a}=n_{0}w^a\), $q^a_0 = h_{0}w^a$ \cite{Van11p}. \re{geqnbal}-\re{geqsbal} and \re{eqnbal}-\re{eqsbal} are the  forms of the same equilibrium fields in the J\"uttner and in the general frames respectively. In the specific equilibrium the J\"uttner, Eckart and Landau-Lifshitz frames coincide, the different choices lead to the same condition: $w_a=0$.

\section{Thermodynamics of relativistic fluids -- out of equilibrium}

In classical non-equilibrium thermodynamics, without internal variables, one assumes that the gradients of the equilibrium fields characterize the deviation from local kinetic equilibrium. In that case the concept of local equilibrium is not modified. The internal variable theories, like the Israel-Stewart theory \cite{Mul69a,Isr76a,IsrSte79a1,IsrSte79a,Ste77a} or GENERIC \cite{Ott98a,Ott05b}, choose a different  characterization: local equilibrium is modified, some {\em formerly dissipative currents appear among the state variables and as a conseqence their contribution may reduce the entropy production.}  
The relativistic theories  revealed that the flow-frame fixing plays a special role in the specification of local equilibrium. It has been an observation of Planck and Einstein, that the momentum density (energy current density) is not purely dissipative and therefore in relativistic theories it has to be taken into account even in local equilibrium \cite{Pla08a,Ein07a}. 

Our starting point is the fundamental inequality of the second law \re{entrcond}. We introduce the following relation of the fields out of equilibrium,  as a generalization of \re{eqrel}: 
\be
S^a+\alpha N^a-\beta_aT^{ab} = \Phi^a.
\ee{neqrel}
With a general  \(\Phi^a\) this relation is valid without any restriction. In a general flow-frame,  \(u^{a}\), we define  the thermostatic pressure as:
\be
p=\frac{u_a \Phi^a}{\beta}.
\ee{tspre}   
Therefore the general form of the potential  \(\Phi^a\) can be written as   
\be
\Phi^a=\beta p(u^a + g^a), \quad \text{where} \quad u^ag_a=0.
\ee{tsphre}   
The parallel and perpendicular components of \re{neqrel} to the flow \(u^a\) are
\ba
s+\alpha n- \beta(h+w_bq^b) &=& 0,\label{neqrel_t}\\
J^a+ \alpha j^a - \beta(q^a + w_b \Pi^{ab}) + \beta p(w^a-g^a) &=& 0^a, 
\ea{neqrel_s}
where \(h=e+p\) and \(\Pi^{ab} = P^{ab}+p\Delta^{ab}\) is the viscous pressure. Then we rewrite the entropy production \re{entrcond} with flow related quantities:
\ba
\Sigma &=& \partial_a S^a +\alpha \partial_aN^a - \beta_a \partial_b T^{ab} = -N^a\partial_a\alpha+T^{ab}\partial_a\beta_b+\partial_a\Phi^a =\nonumber\\
& &\underline{-n\dot \alpha +h\dot\beta+q^a(\beta w_a\dot)+\beta\dot p}+\Pi^{ab}\partial_a\beta_b-j^a \partial_a\alpha + q^a\partial_a \beta +\nonumber\\
&& \beta\dot u_b(q^b-hw^b)+
\beta q^au^b\partial_aw_b + 
p(g^a - w^a)\partial_a \beta + \nonumber\\
&& p\beta\partial^a(g^a-w_a)+g^a\beta \partial_ap.
\ea{etrpr1}
Thermodynamics is taken into account by the following two postulates. 

1) \textit{The underlined part in the above expression with proper time derivatives (total differentials) is zero.}
\be
\beta\frac{d}{dt} p=n\frac{d}{dt} \alpha -h\frac{d}{dt}t\beta -q^a\frac{d}{dt}(\beta w_a).
\ee{GDdif}
This is the relativistic Gibbs-Duhem relation. Considering this together with the vanishing differential of \re{neqrel_t}, we obtain the Gibbs relation \cite{BirVan10a}:
\be
\beta (\d e +w_a\d q^a)=\d s+\alpha\d n.
\ee{Grel}
Based on this result we {conclude} that the entropy has to be given by a functional relationship between the local densities (but certainly including the momentum density \(q^{a}\)), i.e. the proper relativistic and local equation of state is a function \(s(e,q^a,n)\). It has the following partial derivatives:
\be
\left.\pd{s}{e}\right|_{q^a,n} = \beta, \qquad 
\left.\pd{s}{n}\right|_{e,q^a} =-\alpha, \qquad \left.\pd{s}{q^a}\right|_{e,n} = \beta w_a,
\ee{entrpder}
identifying the thermodynamical entropic intensive parameters as being $\beta$, $\alpha =\beta \mu$ and $\beta w^a$. The four-vector $w^a$ is constrained by its orthogonality to the local flow, so it contains independent information on a spatial three-vector only. In isotropic media this degree of freedom is reduced to the length of this vector, $w^2$. In cases containing radiation it appears as a velocity parameter of the Doppler shift \cite{BirVan10a}.

By utilizing the above functional form of the equation of state one derives that the pressure, the intensive parameter associated to mechanical work, satisfies the following four-vector generalized Gibbs-Duhem relation, now written by the traditional differentials:
\be
\beta\partial_a p = n\partial_a \alpha-h\partial_a\beta -q^b\partial_a(\beta w_b).
\ee{GDrel_f}

2) \textit{Our second postulate is \(g^{a}=w^a\).}  By doing so we spell out the fundamental compatibility of non-equilibrium \re{neqrel} with the equilibrium \re{eqrel} definitions of pressure. In this way we treat the non-dissipative part of the thermodynamical potential, and with that the influence of the pressure gradient on the entropy production rate possibly closest to the ideal gas behavior. This is a special matching condition known from kinetic theory ($\delta n =0$, $\delta e=0$): in this case the pressure four-vector $\Phi^a$ is parallel to the thermometer vector $\beta^a$.

Now a short calculation reduces  \re{etrpr1} to a form collecting terms according to the gradients of intensives. A chemical diffusion part is associated to $\partial_a\alpha$, a heat diffusion (Fourier-) part to the gradient of $\beta$, and finally a viscosity term with the symmetric gradient tensor of the full four-vector $\partial_a\beta_b$. We also gain one further term containing the gradient of the difference between $u^a$ and $w^a$. The antisymmetry of the multiplier enforces the antisymmetry of this velocity related gradient, therefore this term we tag as ''vorticity''. We arrive at the following expression:
\ba
\Sigma &=& (nw^a-j^a) \partial_a\alpha + (q^a-hw^a)(\partial_a \beta+\beta \dot u_a) \nonumber\\
&+& (\Pi^{ab}-q^{(a}w^{b)})
\partial_a\beta_b+ q^{[b}w^{a]}\partial_a(\beta (u_b-w_b))\geq 0. 
\ea{etrpr}
Here \(q^{(a}w^{b)}\) and \(q^{[a}w^{b]}\) denotes the symmetric and antisymmetric parts of \(q^{a}w^{b}\) respectively. \re{etrpr} is the entropy production rate without fixing the flow-frame. For a  perfect fluid, characterized by \re{geqnbal}-\re{geqTbal}, the local entropy production is zero. Now it is straightforward to identify thermodynamic fluxes and forces and establish  functional relationships, that are strictly linear in the first approximation\footnote{Since dissipative fluxes are orthogonal to $u^a$, only the projected gradient terms, $\Delta^a_b\partial^b$, constitute thermodynamical forces.}: 
\begin{center}
\resizebox{\linewidth}{!}{
\begin{tabular}{c|c|c|c|c}
       & Diffusive & Thermal & Viscous & Vortical \\ \hline
Fluxes & $nw^a-j^a $ & 
    $q^a-hw^a $ & 
    $\Pi^{ab}-q^{(a}w^{b)}$ &
    $q^{[b}w^{a]}$\\ \hline
Forces &$ \nabla^a\alpha $ &
    $\nabla^a \beta+\beta \dot u^a$ &
    $\Delta^{(bc}\nabla^{a)}\beta_c$ &
    $\Delta^{[bc}\nabla^{a]}(\beta (u_c-w_c))$\\
    \end{tabular}
}\\
\vskip .1cm
{Table 1. Thermodynamic fluxes and forces in a general flow frame}\end{center}

Here \(\nabla_{a} = \Delta_{a}^b\partial_b\). The corresponding linear response relations for isotropic continua are:
\ba
nw^a-j^a &=& D\nabla^a\alpha +\sigma(\nabla^a \beta+\beta \dot u^a), \label{or_diff}\\
q^a-hw^a &=& \sigma\nabla^a\alpha +\lambda(\nabla^a \beta+\beta \dot u^a), \label{or_ther}\\
\Pi^{ab}-q^{(a}w^{b)} &=& \zeta\Delta^{ab}\partial^c\beta_c+2\eta \Delta^{\langle bc}\nabla^{a\rangle}\beta_c,
 \label{or_ms}\\
 q^{[b}w^{a]} &=& \chi \Delta^{[bc}\nabla^{a]}(\beta (u_c-w_c)).
 \ea{or_mas}
Here \(\langle\rangle\) denotes the symmetric traceless part in the bracketed indices, \(\lambda\) is the heat conduction coefficient, \(D\) is the diffusion coefficient, \(\sigma\) is the Soret-Dufour coefficient of thermal diffusion. \(\zeta\) is the bulk viscosity, \(\eta \) is the shear viscosity, and \(\chi\) is the vortical viscosity coefficient.  Because of  the nonnegative entropy production  \re{etrpr} the linear transport coefficients must fulfill the following inequalities:
\be
D\geq 0, \quad \lambda\geq 0, \quad \lambda D-\sigma^2\geq 0, \quad
\zeta \geq 0, \,\, \eta \geq 0,  \,\, \chi>0. 
\ee{or_ineq}
Here the first three inequalities are coupled channel conditions for stability, while the last three are independent ones.

The procedure described here ensures the existence of a  homogeneous flow field as a time independent solution of the equations of motion of the fluid. {\em That is why deviation from local equilibrium is best characterized by gradients of the basic fields in the first approximation}. 

In the following we study some important particular choices for the flow-frame.

\section{Thermometer frame}

The thermometer or J\"uttner frame is the natural choice in kinetic theory calculations. In this case the direction of \(\beta^{a}\) defines the flow-frame similarly to the natural frame in perfect fluids: \(\beta =\sqrt{\|\beta_b\beta^a \|}\) and \(u^a =\beta^a/\beta\). In this section we apply this definition of the flow-frame. Then the  local equilibrium relations are:
\be
s+\alpha n- \beta h = 0, \qquad \d s + \alpha \d n -\beta \d e = 0. 
\ee{leq_N}
The entropy current density, $J^a$ satisfies
\be
J^a + \alpha j^a - \beta q^a= 0^a, 
\ee{eflo_N}
and the entropy production rate fulfills the inequality,
\ba
\Sigma = -j^{a} \partial_a\alpha + q^a(\partial_a \beta+\beta \dot u_a) + 
\beta\Pi^{ab}\partial_au_b\geq 0.
\ea{entrpr_Jf}

This form of the entropy production was derived originally by Eckart restricting to the case  $g^a = w^a =0$. 

Eckart identified the following thermodynamic fluxes and forces  
\begin{center}
\begin{tabular}{c|c|c|c}
       & Diffusive & Thermal & Mechanical\\ \hline
Fluxes & $-j^{a} $ & 
    $q^a$ & 
    $\beta\Pi^{ab}$\\ \hline
Forces &$ \nabla^a\alpha $ &
    $\nabla^a\beta +\beta\dot u^a$ &
    $\Delta^{(bc}\nabla^{a)}u_c$ \\
    \end{tabular}\\
\vskip .1cm
{Table 2. Thermodynamic fluxes and forces by Eckart.}\end{center}

Unfortunately in this case a generic instability occurs, the linear instability of the homogeneous equilibrium, as it was proved by Hiscock and Lindblom in \cite{HisLin85a}. Nonnegative entropy production is established only if considering the basic balance equations (for energy, momentum and further conserved Noether-charges) as constraints. However, by deriving \re{entrpr_Jf} the balance of momentum does not enter the calculations. Therefore the linear relation between the  thermal part of the fluxes and forces with the acceleration term, $\beta\dot{u}^a$, connects changes in these quantities irrespective to the momentum balance equation \re{momc_bal}. A correct treatment of thermodynamic forces and fluxes on the other hand should introduce the momentum balance into the above entropy production formula. A short calculation leads to:
\ba
\Sigma &=& \left(\frac{n}{h}q^a-j^{a}\right) \partial_a\alpha 
-\frac{\beta}{h}q^a\left(\dot q_a + q_a\partial_b u^b + q^b\partial_b u_a+\partial_b\Pi^b_{\;a}\right) \nonumber\\
&+&\beta\Pi^{ab}\partial_au_b\geq 0.
\ea{etrpr_Nf}
This step makes an important difference with respect to stability properties of the homogeneous equilibrium of a fluid. The corresponding thermodynamic fluxes and forces in the J\"uttner frame are 
\begin{center}
\resizebox{\linewidth}{!}{
\begin{tabular}{c|c|c|c}
       & Diffusive & Thermal & Mechanical\\ \hline
Fluxes & $\frac{n}{h}q^a-j^{a} $ & 
    $-\frac{\beta}{h}q^a$ & 
    $\beta\Pi^{ab}$\\ \hline
Forces &$ \nabla^a\alpha $ &
    $X^{a}=\Delta^{ab}\dot q_b + q^a\partial_b u^b+q^b\partial_b u^a+\Delta^{ac}\partial_b\Pi^{b}_c$ &    $\Delta^{(bc}\nabla^{a)}u_c$ \\    
\end{tabular}}\\
\vskip .1cm
{Table 3. Thermodynamic fluxes and forces in J\"uttner frame providing generic stability}\end{center}

Here \(X^a\) is a convenient abbreviation for the thermal force, the thermodynamical force associated to the dissipative current of the heat. Linear transport relations for isotropic continua in the J\"uttner frame can now be easily established:
\ba
\frac{n}{h}q^a-j^{a} &=& D\nabla^a\alpha +\sigma\ X^a, \label{or_Ndiff}\\
-\frac{\beta}{h}q^a &=& \sigma\nabla^a\alpha +\lambda X^{a}, \label{or_Nther}\\
\beta\Pi^{ab} &=& \zeta\Delta^{ab}\partial_cu^c+2\eta \Delta^{\langle b}_{\,\, c}\nabla^{a\rangle}u^c.
\ea{or_Nm}

With this modification the generic stability of the theory in J\"uttner fframe is established:
the heat transfer vector $q^a$ receives a positive relaxation factor, $\beta/h\lambda > 0$.
It is easy to realize that by ignoring viscosity, component diffusion and cross effects,
in homogeneous equilibrium, where all spacelike projected gradients of the velocity field vanish,
the only surviving term in the thermal force is that with the total time derivative of the heat vector:
\be
\lambda X^a =  \lambda \, \Delta^{ab} \dot{q}_b = - \frac{\beta}{h} \, q^a.
\ee{MY_DERIV}
Multiplied by $q^a$ this leads to a relaxation equation for the length of the vector, $Q=-q_aq^a$ as follows
\be
\dot{Q} = - 2\frac{\beta}{h \lambda} \, Q.
\ee{Q2_RELAX}
This means a relaxation towards the  $q^a = 0$ value of the energy current density. 

An important property of these equations is the expected generic statiblity of the homogeneous equilibrium.
Without the detailed calculations (to be shown elswhere) we want to emphasize that the conditions of generic stability are purely thermodynamic. Namely, it is fulfilled whenever the transport coefficients \(\lambda,\tilde\eta\) are nonnegative and the following inequalities for thermodynamic stability i.e. the concavity of the entropy $s(e,n,q^a)$ are satisfied: 
\be
\partial_e T > 0, \qquad \partial_n \frac{\mu}{T} > 0
\qquad \partial_e T \partial_n \frac{\mu}{T} - \left(\frac{\partial_nT}{T}\right)^2 \geq 0. 
\ee{FNS_STABLE}

\section{Other flow-frames}

The other flow frames can be conveniently defined in our general framework. 

The Eckart frame is defined by the direction of the particle current density vector \(u^a=N^a/\sqrt{\|N^bN_b\|}\).  One realizes that in case of dissipative fluids the J\"uttner and Eckart frames do not coincide.

In case of a Landau-Lifshitz frame  the flow field is defined by the direction of the momentum density vector 
\(u^a =u^bT_b^a/\|u^cT_c^d\|\), therefore \(q^{a}=0^a\). 
In case of dissipative fluids the J\"uttner and Landau-Lifshitz frames also do not coincide. However, in the absence of \(q^{a}\), the  thermodynamic relations are similar to the ones in a J\"uttner frame 
\be
s+\alpha n- \beta h = 0, \quad \d s + \alpha \d n -\beta \d e = 0, \quad 
n\d\alpha -h\d\beta -\beta\d p= 0.
\ee{leq_LL}

In principle there are several further possibilities of frame fixing. One of them introduces $w^a = \beta^a/h$.  This choice fixes the velocity field compatible to some kinetic theory calculations \cite{Van11p,VanBir12a}.

Once a choice of the linear response has been made, one can transform the description in one frame to the other. The different transport coefficients are not equivalent, a constant invariant coefficient may become flow-frame dependent in other frames. Wether the primary flow-frame independent choice is preferred or not requires further investigations.

\section{Summary}

Thermodynamic relations in relativistic fluids adhore to flow-frames, while dividing spacial homogeneous changes from the forces enforcing this homogeneity. It is made transparent in the train of thoughts from \re{neqrel} to \re{etrpr1}, where we calculated   entropy production separating comoving time derivatives and spacial gradients. \textit{We have seen, that \(\alpha\), \(\beta^a\) and \(p\) are flow-frame independent.} Then local equilibrium was postulated by the thermodynamic relation \re{Grel},  containing homogeneous thermodynamics. In \cite{BirVan10a}, presenting a different reasoning, we have shown that the different transformation formulas of the relativistic temperature, due to Planck-Einstein, Blanu\u sa-Ott, Landsberg and Doppler, can be unified and reasonably explained in exactly this thermodynamic framework.    

We propose that the thermometer frame, defined in \re{Jflow}, should be a preferred choice. In general \(\beta^a\) can be divided into parts orthogonal and parallel to the flow \(u^a\): \(\beta^a = \beta(u^a+w^a)\), where \(u^aw_a=0\). We have revealed how far this choice differs from the Eckart and Landau-Lifshitz frames. 
There are arguments, that the widely used Landau-Lifshitz frame should be perferred \cite{KosLiu00a,TsuKun13a}. However, these studies do not distinguish the thermometer frame.

The entropy production in a general frame \re{etrpr} helps to recognize 
\begin{itemize}
\item -- that viscous pressure is damps the inhomogeneities in \(\beta^a\), 
\item -- that there are perfect fluids with zero entropy production but \(j^{a}/n=q^a/h=w^a\neq 0\) and \(\Pi^{ab}=hw^aw^b\neq 0\), 
\item  -- there is a vorticity related dissipative term. 
\end{itemize}
Furthermore we have mentioned, that  generic stability is properly derived if the momentum balance constraint is also considered in the calculation of the entropy production \re{entrpr_Jf}. 

In our previous works we have shown further examples of flow-frames. In \cite{VanBir08a,Van09a,Van08a,BirVan10a} the \(w^{a}=q^a/e\) case was explored and in \cite{Van11p} and \cite{VanBir12a} we have analyzed the kinetic theory compatibility and thermodynamics when  \(w^a = q^a/h\). Thermodynamic considerations show, that the coupling of the momentum balance to the entropy production cannot be avoided \cite{Van08a}. It was proven independently  of the Eckart or Landau-Lifshitz frame for \(w^{a}=q^a/e\) in \cite{Van09a}, for the  \(w^a = q^a/h\) case a partial proof was given in \cite{VanBir12a}. 


\section{Acknowledgement}   

The work was supported by the grants Otka K81161, K104260 and TÉT 10-1-2011-0061/ZA-15-2009. The authors thank Etele Moln\'ar for valuable discussions.


\begin{thebibliography}{10}

\bibitem{Ger01m}
R.~Geroch.
\newblock On hyperbolic "theories" of relativistic dissipative fluids.
\newblock 2001.
\newblock arXiv:gr-qc/0103112.

\bibitem{Ger95a}
R.~Geroch.
\newblock Relativistic theories of dissipative fluids.
\newblock {\em Journal of Mathematical Physics}, 36(8):4226--4241, 1995.

\bibitem{GerLin90a}
R.~Geroch and L.~Lindblom.
\newblock Dissipative relativistic fluid theories of divergence type.
\newblock {\em Physical Review D}, 41:1855--1861, 1990.

\bibitem{GerLin91a}
R.~Geroch and L.~Lindblom.
\newblock Causal theories of dissipative relativistic fluids.
\newblock {\em Annals of Physics}, 207:394--416, 1991.

\bibitem{LiuAta86a}
I-S. Liu, I.~M\"uller, and T.~Ruggeri.
\newblock Relativistic thermodynamics of gases.
\newblock {\em Annals of Physics}, 169:191--219, 1986.

\bibitem{PerCal10a}
J.~Peralta-Ramos and E.~Calzetta.
\newblock Divergence-type nonlinear conformal hydrodynamics.
\newblock {\em Physical Review D}, 80:126002, 2010.

\bibitem{ShiEta10a}
Shi Pu, T. Koide, and Dirk~H. Rischke.
\newblock Does stability of relativistic dissipative fluid dynamics imply
  causality?
\newblock {\em Physical Review D}, 81:114039, 2010.

\bibitem{HisLin83a}
W.~A. Hiscock and L.~Lindblom.
\newblock Stability and causality in dissipative relativistic fluids.
\newblock {\em Annals of Physics}, 151:466--496, 1983.

\bibitem{Fic92a}
G.~Fichera.
\newblock Is the {F}ourier theory of heat propagation paradoxical?
\newblock {\em Rediconti del Circolo Matematico di Palermo}, XLI:5--28, 1992.

\bibitem{KosLiu00a}
P.~Kost\"adt and M.~Liu.
\newblock On the causality and stability of the relativistic diffusion
  equation.
\newblock {\em Physical Reviews D}, 62:023003, 2000.

\bibitem{Cim04a}
V.~A. Cimmelli.
\newblock On the causality requirement for diffusive-hyperbolic systems in
  non-equilibrium thermodynamics.
\newblock {\em Journal of Non-Equilibrium Thermodynamics}, 29(2):125--139,
  2004.

\bibitem{VanBir08a}
P.~V\'an and T.~S. B\'\i{}r\'o.
\newblock Relativistic hydrodynamics - causality and stability.
\newblock {\em The European Physical Journal - Special Topics}, 155:201--212,
  2008.

\bibitem{DenAta08a1}
G.~S. Denicol, T.~Kodama, T.~Koide, and Ph. Mota.
\newblock Stability and causality in relativistic dissipative hydrodynamics.
\newblock {\em Journal of Physics G - Nuclear and Particle Physics},
  35(11):115102, 2008.

\bibitem{HisLin85a}
W.~A. Hiscock and L.~Lindblom.
\newblock Generic instabilities in first-order dissipative relativistic fluid
  theories.
\newblock {\em Physical Review D}, 31(4):725--733, 1985.

\bibitem{TsuKun08a}
K.~Tsumura and T.~Kunihiro.
\newblock Stable first-order particle-frame relativistic hydrodynamics for
  dissipative systems.
\newblock {\em Physics Letters B}, 668(5):425--428, 2008.

\bibitem{TsuKun11a}
K.~Tsumura and T.~Kunihiro.
\newblock First-principle derivation of stable first-order generic-frame
  relativistic dissipative hydrodynamic equations from kinetic theory by
  renormalization-group method.
\newblock {\em Progress of Theoretical Physics}, 126(5):761--809, 2011.

\bibitem{Van09a}
P.~V\'an.
\newblock Generic stability of dissipative non-relativistic and relativistic
  fluids.
\newblock {\em Journal of Statistical Mechanics: Theory and Experiment}, page
  P02054, 2009.

\bibitem{VanBir12a}
P.~V\'an and T.S. Bir\'o.
\newblock First order and generic stable relativistic dissipative
  hydrodynamics.
\newblock {\em Physics Letters B}, 709(1-2):106--110, 2012.

\bibitem{GarEta09a}
A.~L. Garcia-Perciante, L.~S. Garcia-Colin, and A.~Sandoval-Villalbazo.
\newblock On the nature of the so-called generic instabilities in dissipative
  relativistic hydrodynamics.
\newblock {\em General Relativity and Gravitation}, 41(7):1645--1654, 2009.

\bibitem{Osa10a}
T.~Osada.
\newblock Relativistic hydrodynamical model in the presence of long-range
  correlations.
\newblock {\em Physical Review C}, 81:024907, 2010.

\bibitem{Osa12a}
T.~Osada.
\newblock Modification of Eckart theory of relativistic dissipative fluid
  dynamics by introducing extended matching conditions.
\newblock {\em Physical Review C}, 85:014906, 2012.

\bibitem{GroAta80b}
S.R. de~Groot, W.~A. van Leeuwen, and Ch.~G. van Weert.
\newblock {\em Relativistic Kinetic Theory}.
\newblock North Holland, Amsterdam, 1980.

\bibitem{Cse94b}
Csernai L.
\newblock {\em Introduction to relativistic heavy ion physics}.
\newblock John Wiley and Sons, Chicester-etc., 1994.

\bibitem{Van11p}
P.~V\'an.
\newblock Kinetic equilibrium and relativistic thermodynamics.
\newblock {\em EPJ WEB of Conferences}, 13:07004, 2011.

\bibitem{IsrSte79a1}
W.~Israel and J.~M. Stewart.
\newblock On transient relativistic thermodynamics and kinetic theory.
\newblock {\em Annals of Physics}, 118:341--372, 1979.

\bibitem{MonHir09a}
A~Monnai and T.~Hirano.
\newblock Effects of bulk viscosity at freezeout.
\newblock {\em Physical Review C}, 80:054906, 2009.

\bibitem{Eck40a3}
Carl Eckart.
\newblock The thermodynamics of irreversible processes, {III}. {R}elativistic
  theory of the simple fluid.
\newblock {\em Physical Review}, 58:919--924, 1940.

\bibitem{LanLif59b}
L.~D. Landau and E.~M. Lifshitz.
\newblock {\em Fluid mechanics}.
\newblock Pergamon Press, London, 1959.

\bibitem{Far894a}
Gy. Farkas.
\newblock A {F}ourier-f\'ele mechanikai elv alkalmaz\'asai.
\newblock {\em Mathematikai \'es Term\'eszettudom\'anyi \'Ertes\'\i{}t\H{o}},
  12:457--472, 1894.
\newblock in Hungarian.

\bibitem{Min896b}
H.~Minkowski.
\newblock {\em Geometrie der Zahlen}.
\newblock Teubner, Leipzig und Berlin, 1896.

\bibitem{Sch98b}
A.~Schriver.
\newblock {\em Theory of linear and integer programming}.
\newblock John Wiley and Sons, Chicester-etc.., 1998.

\bibitem{TsuKun12a}
K.~Tsumura and T.~Kunihiro.
\newblock Derivation of relativistic hydrodynamic equations consistent with
  relativistic {B}oltzmann equation by renormalization-group method.
\newblock {\em The European Physical Journal A}, 48:162, 2012.

\bibitem{Mul69a}
I.~M\"uller.
\newblock Toward relativistic thermodynamics.
\newblock {\em Archive for Rational Mechanics and Analysis}, 34(4):259--282,
  1969.

\bibitem{Isr76a}
W.~Israel.
\newblock Nonstationary irreversible thermodynamics - causal relativistic
  theory.
\newblock {\em Annals of Physics}, 100(1-2):310--331, 1976.

\bibitem{IsrSte79a}
W.~Israel and J.~M. Stewart.
\newblock On transient relativistic thermodynamics and kinetic theory. {II}.
\newblock {\em Proceedings of Royal Society London A}, 365:43--52, 1979.

\bibitem{Ste77a}
J.~M. Stewart.
\newblock On transient relativistic thermodynamics and kinetic theory.
\newblock {\em Proceedings of Royal Society London A}, 357:59--75, 1977.

\bibitem{Ott98a}
H.~C. \"Ottinger.
\newblock On the structural compatibility of a general formalism for nonequilibrium dynamics with special relativity.
\newblock {\em Physica A}, 259:24-42, 1998.

\bibitem{Ott05b}
H.~C. \"Ottinger.
\newblock {\em Beyond equilibrium thermodynamics}.
\newblock Wiley-Interscience, 2005.

\bibitem{Pla08a}
M.~Planck.
\newblock Zur {D}ynamik bewegter {S}ysteme.
\newblock {\em Annalen der Physik}, 331(6):1--34, 1908.

\bibitem{Ein07a}
A.~Einstein.
\newblock {\"U}ber das {R}elativit\"atsprinzip und die aus demselben gezogenen
  {F}olgerungen.
\newblock {\em Jahrbuch der Radioaktivit\"at und Elektronik}, 4:411--462, 1907.

\bibitem{BirVan10a}
T.~S. B\'ir\'o and P.~V\'an.
\newblock About the temperature of moving bodies.
\newblock {\em EPL}, 89:30001, 2010.

\bibitem{TsuKun13a}
K.~Tsumura and T.~Kunihiro.
\newblock Uniqueness of Landau-Lifshitz energy frame in relativistic
  dissipative hydrodynamics.
\newblock {\em Physical Review E}, 87:053008, 2013.

\bibitem{Van08a}
P.~V\'an.
\newblock Internal energy in dissipative relativistic fluids.
\newblock {\em Journal of Mechanics of Materials and Structures},
  3(6):1161--1169, 2008.

\end{thebibliography}
\end{document}